\documentclass[12pt]{article}

\usepackage{url}

\usepackage[utf8]{inputenc} 
\usepackage[english]{babel} 
\usepackage{siunitx} 
\usepackage{graphicx} 
\usepackage{csquotes} 

\usepackage[backend=biber,style=ieee]{biblatex}
\AtBeginBibliography{\footnotesize}

\usepackage{xcolor}

\newcommand{\SimPyLink}{https://simpy.readthedocs.io/}
\newcommand{\SimPyLinkFootnoteText}{SimPy: \SimPyLink}

\providecommand{\keywords}[1]
{
  \small	
  \textbf{\textit{Keywords---}} #1
}

\makeatletter
\def\blfootnote{\gdef\@thefnmark{}\@footnotetext}
\makeatother

\begin{document}

\baselineskip 12pt

\begin{center}
\textbf{\Large {Keeping up with the bits: tracking physical layer latency in millimeter-wave Wi\babelhyphen{nobreak}Fi networks}} \\
\vspace{5mm}
\begin{tabular}{cr}  
Alexander Marinšek & Liesbet Van der Perre\\[1.2\baselineskip]

\multicolumn{2}{c}{KU Leuven} \\
\multicolumn{2}{c}{ESAT-WaveCore} \\
\multicolumn{2}{c}{Ghent Technology Campus, 9000 Ghent, Belgium} \\[1.2\baselineskip]
\multicolumn{2}{c}{\texttt{alexander.marinsek@kuleuven.be}} \\
\end{tabular} \end{center}

\vspace{0.5em}

\blfootnote{This work has been accepted to the 41$^{st}$ WIC Symposium On Information Theory and Signal Processing in the Benelux (SITB 2021).}

\begin{abstract}
\noindent The wireless communications landscape is anticipated to offer new service levels following the introduction of the millimeter\babelhyphen{nobreak}wave (mmWave) spectrum to consumer electronics. With their broad bandwidths and corresponding multi\babelhyphen{nobreak}Gbps data rates, these mmWaves are a perfect fit for data hungry applications, such as streaming video to extended reality devices. However, the latter are also bound by maximal latency constraints as low as 1 ms. Understanding where such minuscule time delays lurk requires a close-up study of individual layers in the network stack. Starting from the bottom up, the present work describes an endeavor at uncloaking the origins of physical layer (PHY) latency in mmWave Wi\babelhyphen{nobreak}Fi networks. It proposes a newly designed simulation framework and sheds light on how any conventional laboratory can be turned into a virtual experiment setting, speeding up computation. A case study based on the IEEE 802.11ad standard demonstrates the framework's ability to track packet latency at the PHY-level and identify individual bottlenecks. In particular, it evaluates the impact of the number of LDPC decoding iterations on latency in short transmission sequences.
\end{abstract}


\section{Introduction}

A growing number of latency sensitive applications are imposing ever stricter end-to-end (E2E) time delay constraints on wireless networks. The use cases range from more traditional ones such as voice over IP (VoIP) \cite{kassim_performance_2017}, to future connected vehicles, industry 4.0, and extended reality. The applications contained therein require extremely low network time delays. For example, coordinated driving \cite{kanavos_v2x_2021}, cooperative robots \cite{5g_alliance_for_connected_industries_and_automation_5g_2019_custom}, and immersive media \cite{lincoln_motion_2016} may all impose E2E latency constraints as low as 1 ms. Moreover, they are starting to increasingly rely on computation offloading \cite{yousefpour_all_2019}, giving rise to their second requirement: high data rates.

Within the broad range of mmWave frequencies, several gigahertz of bandwidth, centered around 60 GHz, have been allocated to unlicensed communications. One of the standards operating in the 60 GHz band is the IEEE 802.11ad mmwave Wi\babelhyphen{nobreak}Fi \cite{ieee_computer_society_directional_2016_custom}, also referred to as WiGig. Supporting 64QAM and high LDPC code rates, it manages to attain data rates of up to 8.085 Gbps while conforming to the 1.76 GHz channel bandwidth limitations. 
This reduces the transmission times of even the longest physical layer (PHY) payloads (262 KB) to well below 1 ms; however, the finite transfer rate of the wireless network is not the only factor determining E2E latency. 

The IEEE 802.11ad PHY features several components, every one of which may either add a small delay to the propagation of data, or in some cases, become a bottleneck. For  both  analysis  purposes  and  the  conception  of  adequate latency mitigation strategies and solutions, a simulation framework has been developed.

\subsection{State of the art simulation frameworks}

Among the existing simulations tools, ns-3 is one of the most widespread publicly available network simulators. It is backed by a lively community, thanks to which it also supports the IEEE 802.11ad standard. Described in \cite{assasa_implementation_2016}, the WiGig ns-3 model features a detailed MAC layer on top of an abstract PHY implementation. It is a versatile tool for higher-level studies, concerning beamforming and fast session transfer among others. It is also able to keep track of transmission delays and it can induce data corruption by leveraging one of the underlying PHY error models. However, it does not cover the PHY components in greater detail, meaning their contribution to latency is unknown.

A substitute for ns-3 that offers more control over the PHY is the set of tools contained within MATLAB and its Communications Toolbox. These offer direct access to PHY component implementations, such as the LDPC codec. The visualization and manual analysis of intermediate results is also simplified with the use of MATLAB's built-in debugging tools. While the toolbox offers control over individual components, some effort was made in the past to encapsulate them in an IEEE 802.11ad PHY simulation framework \cite{blumenstein_ieee_2019}. The framework resides in an open online repository and its functionality is verified against measurement data; yet, it does not support PHY latency tracking.

Another endeavor to provide reliable WiGig simulations was presented in \cite{cordeiro_ieee_2010}, where the OPNET network simulator served as the foundation for the IEEE 802.11ad model. While the authors explicitly stated the timing results obtained during simulation, the values ranging up to 5 ms include MAC layer time delays as latency accumulation within the PHY remains unaddressed.

\subsection{Contribution}

With the progressively stricter latency requirements of future real-time applications, gaining better insight into latency accumulation across the network stack is growing in importance. As the foundation of the network stack, the PHY is also the first layer contributing to latency accumulation. The present work outlines the design principles of a fine-grained simulation framework for tracking latency in the IEEE 802.11ad PHY. Furthermore, it presents a potential usage of the framework for evaluating the impact of LDPC decoding on transmission latency. The contributions are, therefore, two-fold:

\begin{itemize}
\item Design of an open source IEEE 802.11ad PHY latency simulation framework in Python, made publicly accessible through an online repository \cite{alexander_2021_4749534}.
\item Evaluation of the contribution of iterative LDPC decoding to the total latency during the transmission of short payloads (100 KB).
\end{itemize}

\noindent The work is structured as follows. Section \ref{sec:simulation_framework} describes the working principles of the simulation framework and presents a multi-processing approach to reducing execution times both on the local machine and by using otherwise idle remote hardware. Employing the framework on an example, section \ref{sec:simulation_frmework_usage_examples} demonstrates latency tracking in practice and evaluates the delays induced by the LDPC decoder. Finally, section \ref{sec:conclusion} summarizes the main findings and outlines future research prospects.

\section{Simulation framework} \label{sec:simulation_framework}

The key features associated with PHY latency accumulation are finite transmitter (TX) data rate and delays within the receiver (RX) digital baseband (DBB). Simulating data propagation through individual components, the simulation framework is not bound to a particular RX DBB implementation. Assuming basic programming knowledge, the components can be added, removed, or placed at different positions. Moreover, their latency performance figures are freely adjustable. The RX DBB components are represented by individual blocks and buffers within the simulation framework, the inner workings and invocation of which are outlined in the following two sections.

\begin{figure}[h]
	\centering
	\includegraphics[width=\textwidth]{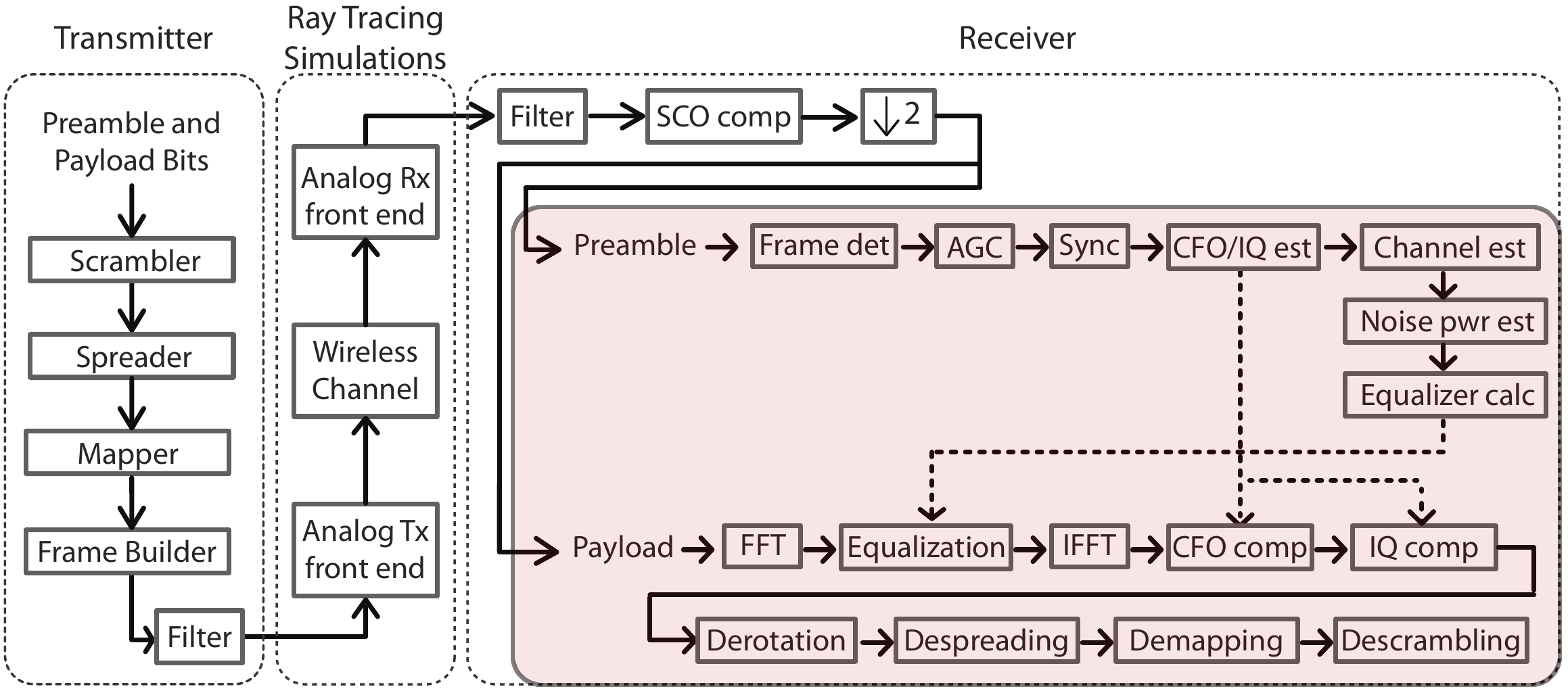}
	\caption{\label{figure:rx_dbb}Overview of the simulated RX DBB (highlighted in red) within the PHY, as presented in \cite{genc_60_2012}.}
\end{figure}

In providing an illustrative example, the present work implements an RX DBB that follows a particular design pattern. It assumes frequency domain equalization and, with minor exceptions, the RX DBB is loosely based on the one shown in Figure \ref{figure:rx_dbb}. 

\subsection{Buffers, blocks, and events}

\begin{figure}[ht]
	\centering
	\includegraphics[width=7cm]{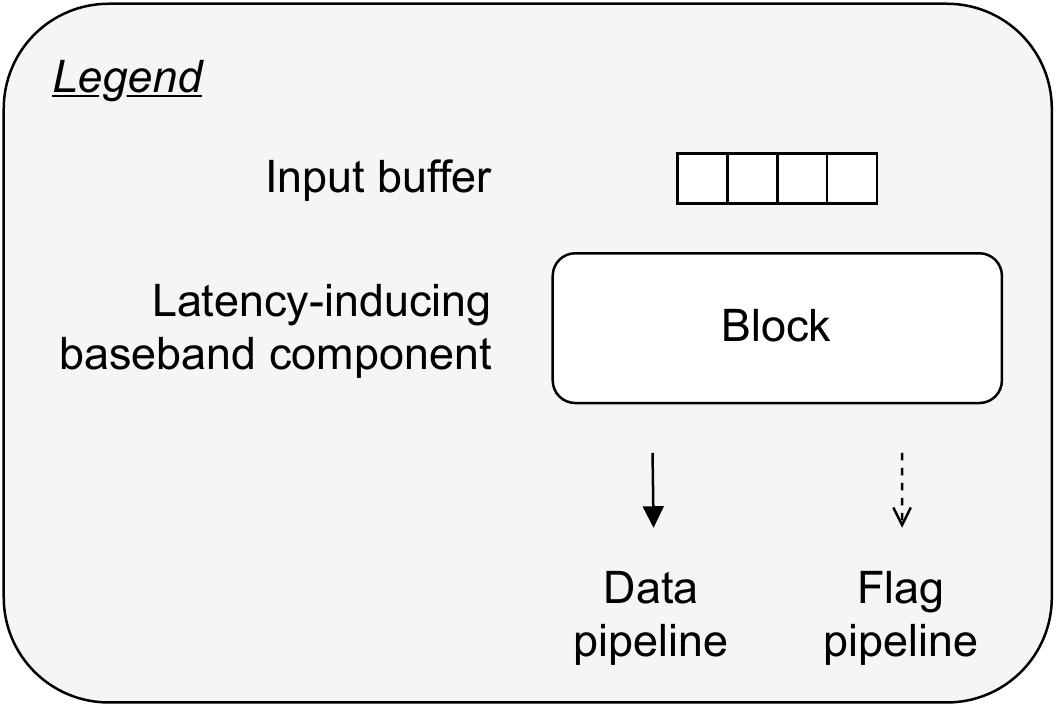}
	\caption{\label{figure:block_legend}Illustration of a block, representing an individual RX DBB component.}
\end{figure}

The propagation of information through the RX DBB takes the form of data passing between the input buffers of individual blocks (components). If the block fails to process the data in time, then they start piling up in its input buffer. Several blocks also feature a second input buffer. Its function is to halt the block from processing the data before being allowed to do so. For example, the equalizer can only start processing the input data once channel estimation has concluded. Corresponding to Figure \ref{figure:rx_dbb} and Figure \ref{figure:block_legend}, solid lines represent pathways leading to input buffers, while dashed lines show the propagation of flag inputs.

Every time data is passed to a buffer, an event is triggered. The simulation framework stores the event type (put, get, request) and the time at which it occurred. Timekeeping is achieved using the SimPy\footnote{\SimPyLinkFootnoteText} Python package. The latter steps through simulated time at a sub-nanosecond resolution and enables time-based data propagation. The pace at which information is passed through the RX DBB is dictated by the performance figures of individual components, sourced from state of the art literature. These are covered in the corresponding module in the simulation framework's online repository.

\subsection{Reducing simulation time}

\begin{figure}[h]
	\centering
	\includegraphics[width=10cm]{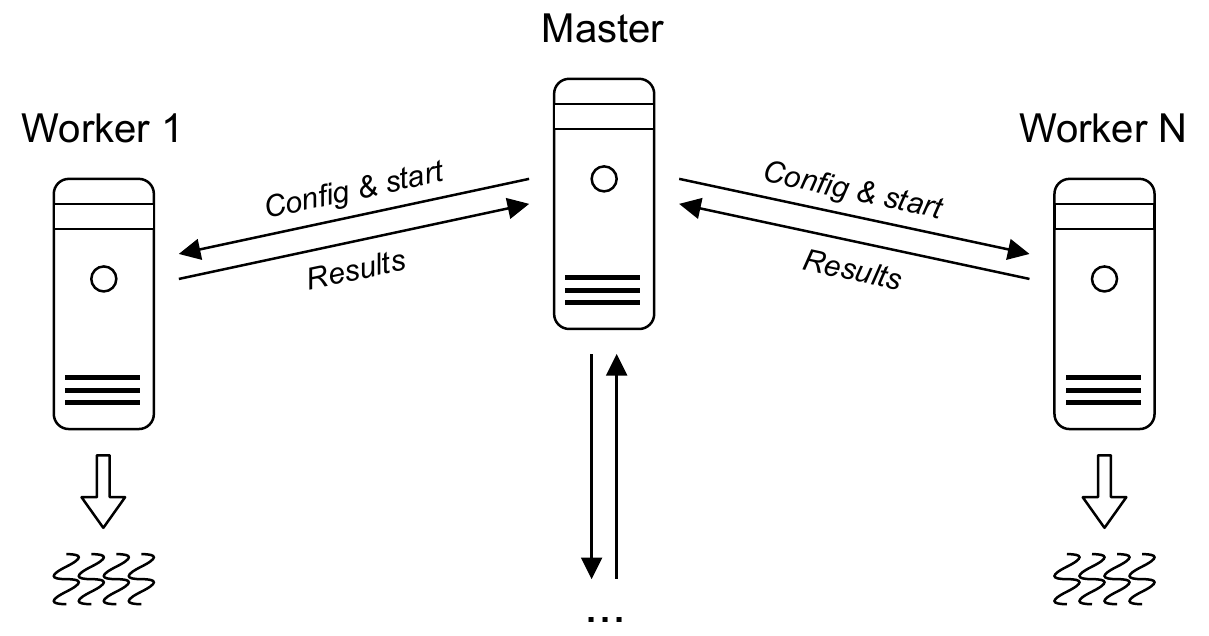}
	\caption{\label{figure:parallel_processing}Example of applying multiprocessing and manual distributed computing for shorter execution times.}
\end{figure}

Executing simulations for multiple payload sizes at different modulation and coding scheme (MCS) settings can present a time consuming task. Especially if only one core on the host machine is employed. Bridging such bottlenecks is often achieved using multithreading. However, being based on Python and tightly connected to SimPy, the simulation framework must instead rely on multiprocessing for parallel computing. Upon starting, the simulation is automatically split into concurrent processes, and as they execute, new ones are spawned based on the simulated parameter combinations. The size of the process pool is manually adjustable. As hinted by figure \ref{figure:parallel_processing}, the execution time is further reduced when multiple machines are used, such as otherwise idle computers in a classroom. 

While process spawning is built into the framework, distributed computing must be set up manually. A typical workflow is to clone the simulation framework to all of the processing machines, configuring the input parameters (payload length, MCS), and starting the simulation in the background. Once finished, the results are manually aggregated before the data analysis is carried out. This is somewhat cumbersome compared to using purpose made distributed computing frameworks, yet, it reduces the number of dependencies and aids overall simplicity of implementation.

\section{Simulation framework usage examples} 
\label{sec:simulation_frmework_usage_examples}

The simulation framework includes detailed event logging in the background. It stores the exact time of every single block buffer input or output operation. Upon simulation execution, the logs are stored on the host machined, together with their corresponding metadata.

\subsection{Latency accumulation in the physical layer}

\begin{figure}[h]
	\centering
	\includegraphics[width=\textwidth]{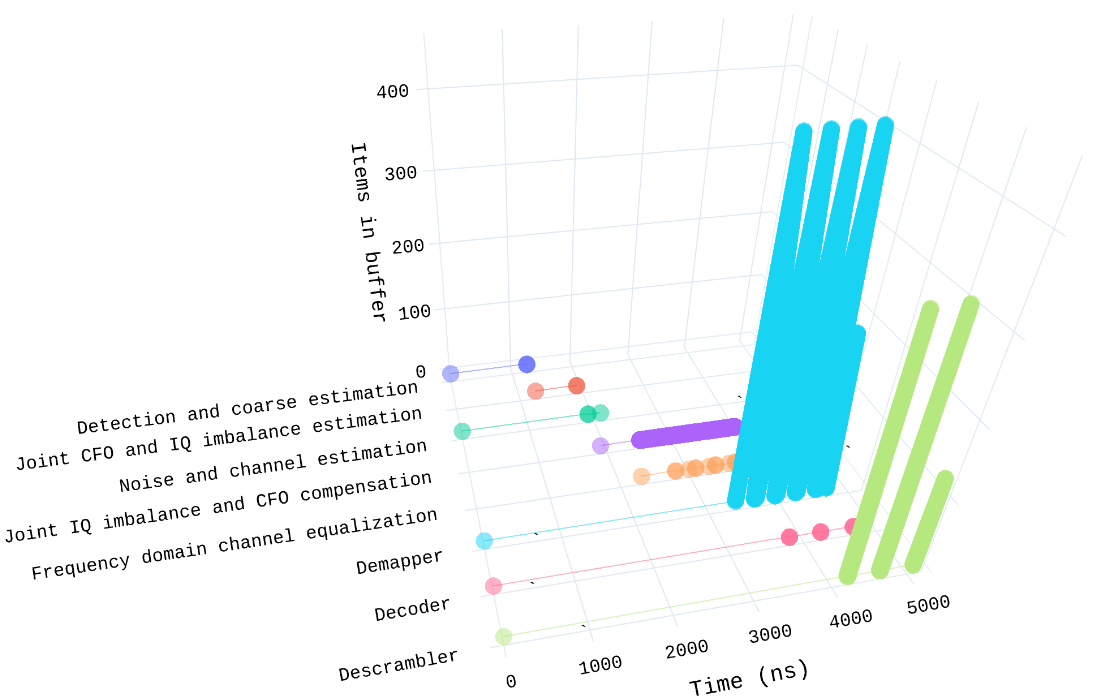}
	\caption{\label{figure:event_overview}Example of a packet bearing 100 B of payload being transmitted using MCS 2.0 and setting the LDPC decoder iteration count to 10. The colored spheres represent buffer events, among which the first one is always the initial data request. Where offset from zero (e.g. Joint CFO and IQ imbalance estimation), there is an additional flag buffer, temporarily preventing the component from processing the incoming data.}
\end{figure}

Illustrating the buffer events during a single transmission, Figure \ref{figure:event_overview} analyzes how data propagate through the RX DBB and where they spend the most time. Observing buffer occupancy, there are two components where data are seemingly piling up: the demapper and the descrambler. This sudden increase in the item count is caused by the equalizer and decoder both outputting entire data blocks. These are then processed as individual symbols and bits by the demapper and descrambler. However, since the descrambler processes all of the data before the arrival of a new dataword, it is not a bottleneck. The same applies to the demapper, which, on the other hand, barely manages to keep up with the incoming data. Mildly reducing its throughput would already make it a bottleneck.

Depending on the transmission parameters (MCS, payload length) and RX DBB configuration (component performance figures), the overview in Figure \ref{figure:event_overview} might take an entirely different form.

\subsection{Evaluating iterative LDPC decoder latencies}

\begin{figure}[h]
	\centering
	\includegraphics[width=\textwidth]{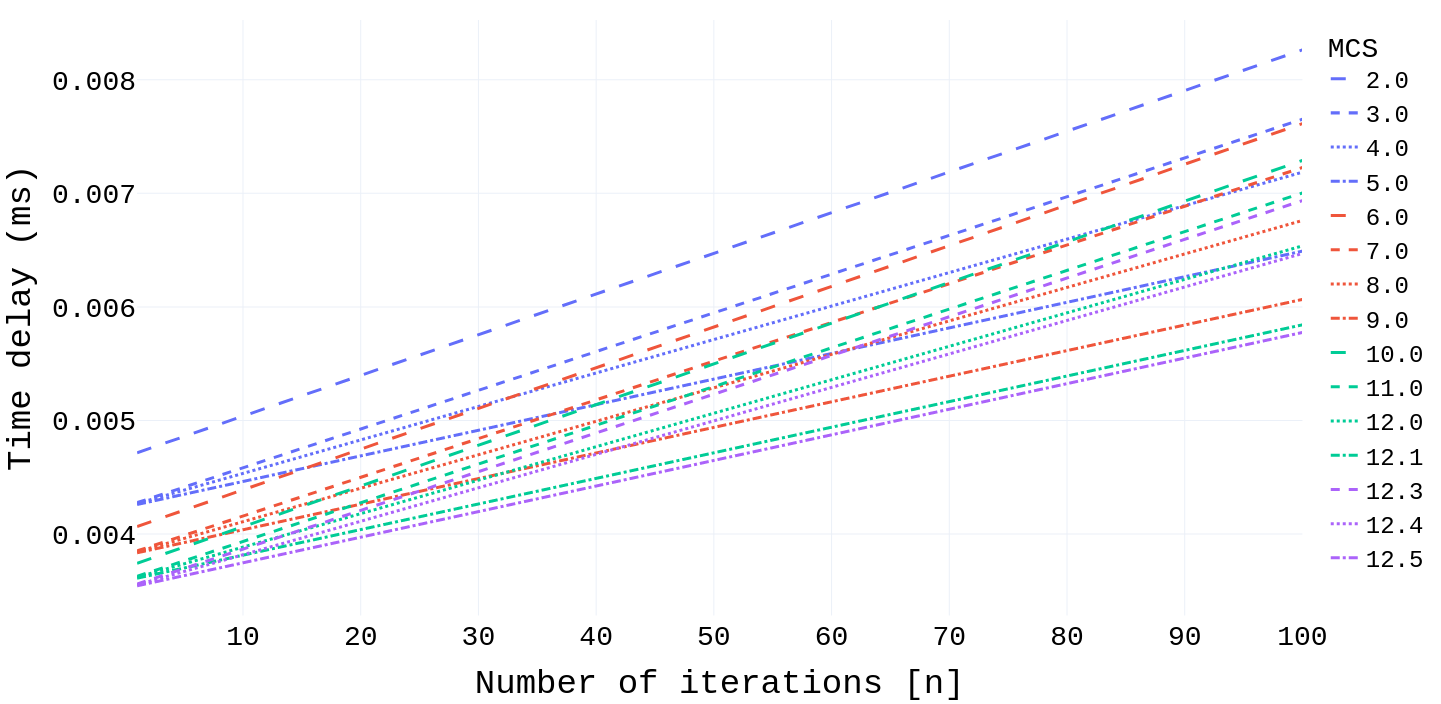}
	\caption{\label{figure:LDPC_decoding}Time delay incurred in the PHY during the transmission and reception of PHY frames bearing 100 B payloads, dependent on the number of LDPC decoding iterations. Line colors represent modulation rate (BPSK, QPSK, 16QAM, 64QAM), while line types stand for different code rates (0.5, 0.625, 0.75, 0.8125).}
\end{figure}

A practical example, where the simulation framework can provide valuable insight, is studying the effects of iterative decoding on the overall latency of the PHY. This is made possible by associating a time delay function with the decoder, dependent on the code rate and number of iterations. The latency results for short transmission sequences are summarized in Figure~\ref{figure:LDPC_decoding}. For each modulation rate, the values obtained using code rate~0.5 feature the highest latency since they spawn the most overhead. Furthermore, the 0.5 and 0.625 code rate lines exhibit a similar slope because of the decoder defining a similar number of needed processing cycles for either of the two code rates \cite{li_processor_2013}. The remaining two code rates, 0.75 and 0.8125, reduce the processing complexity and, therefore, the time delays diverge with the increasing number of iterations.

From another perspective, the topmost and bottom-most lines are the constraints of a new latency region. It outlines the range of expected transmission time delays, regardless of the selected MCS, number of decoder iterations, or both of them.

\section{Conclusion} \label{sec:conclusion}

The present work describes the design and usage of an IEEE 802.11ad PHY simulation framework, intended for PHY time delay analysis and the exploration of latency mitigation strategies to meet strict latency requirements. The core principles of how PHY components are reflected in the simulation framework blocks are outlined, and an approach to reducing execution times using parallel processing is discussed. The results demonstrate the simplicity of tracking data propagation through the PHY components and identifying potential bottlenecks. Finally, the evaluation of LDPC decoder latencies for small payload sizes highlights how the framework can be used in specific studies.

While the presented simulation framework manages to keep track of PHY time delays, it does not address data integrity. Since transmission errors are expected, the next step should focus on associating latency with other performance metrics, such as the bit error rate (BER), and evaluating the trade-offs between the two. Future work can also consider studying specific low-latency applications and the time delays of transmission sequences typically associated with them.

\section{Acknowledgement}

\vspace{1em}
\noindent \begin{minipage}{0.15\textwidth}
\includegraphics[width=0.9\textwidth]{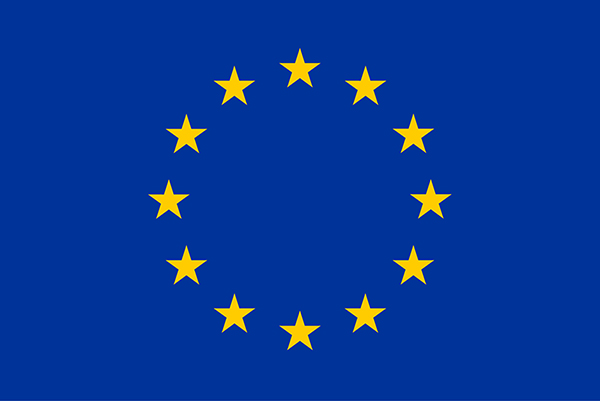}
\end{minipage} \hfill
\begin{minipage}{0.85\textwidth}
 	This project has received funding from the European Union’s Horizon 2020 research and innovation programme under grant agreement No~861222 (MINTS MSCA-ITN). 
\end{minipage}
\vspace{1em}

The authors also wish to acknowledge Gilles Callebaut for providing support during different parts of debugging.

{\footnotesize \printbibliography}

\end{document}